\documentclass[prl,twocolumn]{revtex4}
\usepackage{graphicx}
\usepackage{amsmath}
\usepackage{amssymb}
\usepackage{color}

\def\mnras{MNRAS}
\def\apjl{ApJ Letters}
\def\apjs{ApJ Supplements}

\begin{document}
\title{Pulsar Timing Array Based Search for Supermassive Black Hole Binaries \\ 
in the Square Kilometer Array Era}
\date{\today}
\author{Yan Wang}
\email{ywang12@hust.edu.cn}
\affiliation{School of Physics, Huazhong University of Science and Technology, \\
1037 Luoyu Road,  Wuhan, Hubei Province 430074, China}
\altaffiliation{MOE Key Laboratory of Fundamental Physical Quantities Measurements,
1037 Luoyu Road, Wuhan, Hubei Province 430074, China}
 \author{Soumya D. Mohanty}
\email{soumya.mohanty@utrgv.edu}
\affiliation{Department of Physics, The University of Texas Rio Grande Valley, \\
One West University Boulevard, Brownsville, TX 78520, USA}
\altaffiliation {Center for Gravitational Wave Astronomy,  
The University of Texas Rio Grande Valley,
One West University Boulevard, Brownsville, TX 78520, USA}

\begin{abstract}
The advent of next generation radio telescope facilities, such 
as the Square Kilometer Array (SKA), will usher in an era 
where a Pulsar Timing Array (PTA) based search for gravitational waves (GWs) 
will be able to use hundreds of well timed millisecond
pulsars rather than the few dozens in existing PTAs. 
A realistic assessment of the performance of such an extremely large PTA must 
take into account the data analysis challenge posed by an
exponential increase in the parameter space volume due to
the large number of so-called pulsar phase parameters.
We address this problem and present such an assessment
for isolated supermassive black hole binary (SMBHB) searches using a SKA era PTA
containing $10^3$ pulsars.
We find that an all-sky search will be able to confidently detect
non-evolving sources with redshifted chirp mass 
of $10^{10}$ $M_\odot$  out to a redshift of about $28$ (corresponding to a rest-frame chirp mass of $3.4\times 10^{8}$ $M_\odot$). 
We discuss the important implications that the
large distance reach of a SKA era PTA
has on GW observations from 
optically identified SMBHB candidates. 
If no SMBHB detections occur, 
a highly unlikely scenario in the light of our results, 
the sky-averaged upper limit on strain amplitude will be improved by 
about three orders of magnitude over existing limits.

\end{abstract}

\pacs{abcd}

\maketitle

\paragraph{Introduction --}
Several major efforts are progressing in parallel to open the gravitational wave (GW) window 
in astronomy across a wide range of frequencies. Success has 
been achieved in the high-frequency  band 
($ \sim 10 - 1000$~Hz) with the landmark detection of signals from two binary 
black hole mergers by the Advanced 
Laser Interferometer Gravitational-Wave Observatory (aLIGO) \citep{2016PhRvL.116f1102A, PhysRevLett.116.241103}.
Space-based detectors~\citep{2013arXiv1305.5720C,2009JPhCS.154a2040S,2016CQGra..33c5010L},
for scanning the  $\sim 10^{-4} - 1$~Hz band are in various stages of planning.  
Sensitivities of Pulsar Timing Array (PTA) based GW searches in the 
$\sim 10^{-9} - 10^{-6}$~Hz band continue to improve~\cite{2015Sci...349.1522S, 
2015MNRAS.453.2576L, 2016ApJ...821...13A, 2016MNRAS.458.1267V, 
2014ApJ...794..141A,  2014MNRAS.444.3709Z, 2016MNRAS.455.1665B}.

PTA based GW astronomy will experience a sea change when 
next generation radio telescopes with larger collecting areas and better backend 
systems, such as FAST \citep{2014arXiv1407.0435H} and SKA \citep{2009A&A...493.1161S},
start observations. Simulations based on 
pulsar population models
predict that up to 14000 canonical 
and 
6000 millisecond pulsars (MSPs) can be discovered by SKA \citep{2009A&A...493.1161S}.  
Due to their high intrinsic rotational stability, combined with the improved sensitivity of
SKA, a timing uncertainty of  $< 100$~ns \cite{2004NewAR..48..993K,2010arXiv1004.3602M} is likely
for a substantial fraction of the MSPs. 

The most promising class of GW sources for PTAs is that of 
Supermassive Black Hole Binaries (SMBHBs).
While the number of optically identified SMBHB candidates now ranges in 
the hundreds \cite{2011ApJ...738...20T, 2012ApJS..201...23E,
2015MNRAS.453.1562G, 2016MNRAS.463.2145C}, 
the only unambiguous confirmation of the true nature of 
a candidate is its GW signal. 
If the constraints \citep{2015Sci...349.1522S} on models 
of the unresolved SMBHB population  \citep{2009MNRAS.394.2255S} 
continue to improve in the absence of a detection of the associated 
stochastic signal -- implying a sparser distribution of sources -- the search 
for isolated sources becomes increasingly important.

We  carry out a quantitative 
assessment of the performance one can expect for isolated SMBHB searches 
with an extremely large SKA era PTA containing $10^3$ pulsars.
In order to make the assessment realistic,  the exponential 
growth in the volume of the parameter space defining a GW signal must be taken into account.
This happens because, as explained later, every pulsar in the array introduces a so-called 
pulsar phase parameter whose value is not known {\it a priori}.
This problem is addressed in our analysis by using the algorithm proposed in~\cite{2015ApJ...815..125W}.

\paragraph{Preliminaries --}
Let  $d^I(t)$ denote the timing residual from the $I^{\rm th}$ pulsar, obtained by 
subtracting a fiducial timing model from the recorded pulse arrival times.
The data from an $N$ pulsar PTA can be expressed as \citep{2016MNRAS.461.1317Z}, 
\begin{eqnarray}
\textbf{d}(t) &=& \textbf{A} \Delta\textbf{h}(t) + \textbf{n}(t)   
\label{datalinear}\;.
\end{eqnarray}
Here, $\textbf{d}(t)$ is the column vector whose $I^{\rm th}$ element is $d^I(t)$, and
$\textbf{n}(t)$ is the corresponding column vector of noise in the observations. The 
$I^{\rm th}$ row of the $N$-by-2 response matrix $\textbf{A}$
is comprised of the antenna pattern 
functions $F^{I}_+(\alpha,\delta)$ and $F^{I}_\times(\alpha,\delta)$ 
(their functional forms can be found in \cite{2014ApJ...795...96W}),
with $\alpha$ and $\delta$ being the Right Ascension (RA) and Declination (DEC) of the GW source.
 $\Delta\textbf{h}(t) = (\Delta h_+(t), \Delta h_\times(t))^T$,
 \begin{eqnarray}
\Delta h_{+,\times} (t) & = & h_{+,\times}(t)  - h_{+,\times}(t - \tau_I(\alpha,\delta))\,.
\label{eq:pulsarterm}
\end{eqnarray}
The last term in Eq.~\ref{eq:pulsarterm} is 
called the
{\it pulsar term}. 
The time delay $\tau_I(\alpha,\delta)$ 
depends on the Earth-pulsar distance and the
direction of the source relative to the line of sight to the pulsar. 

\begin{figure}
\centerline{\includegraphics[scale=0.46]{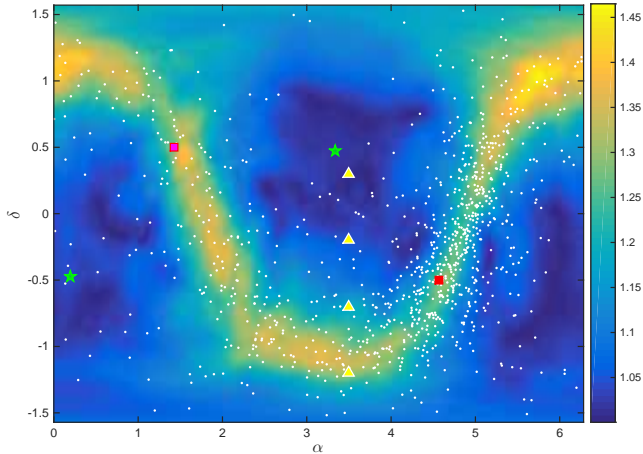}}
\caption{The condition number of the response matrix $\textbf{A}$ as a 
function of RA ($\alpha$) and DEC ($\delta$), both 
expressed in radians. The 
dots show the locations of the MSPs constituting the simulated SKA era PTA used
in this paper. The stars show the Galactic poles (North on top) and the squares show the
Galactic Center (right) and anti-center (left). From top to bottom, the triangles denote the four source
locations A, B, C, and D respectively that are used in the simulations. 
The condition numbers corresponding to these locations in the same order are: 1.0139,  1.0486,  1.1832, and 1.3159.
}
\label{fig:Acond}
\end{figure}

The condition number of  $\textbf{A}$, shown in Fig.~\ref{fig:Acond},
determines the degree of ill-posedness
inherent in the inverse problem \citep{1996..book.....Engl} of estimating signal
parameters from the data. 


\paragraph{Simulated SKA era PTA --}
We construct a realistic SKA era PTA using the simulated 
pulsar catalog in \citep{2009A&A...493.1161S}
and selecting $10^3$  MSPs within 3 kpc from us. 
Fig.~\ref{fig:Acond} shows the locations of the simulated MSPs.

 
We generate data realizations using a uniform cadence for simplicity. 
It is set to two weeks, in order 
to match the typical cadence used in current PTAs.
The span of the simulated timing residuals is 5 years. 
Noise realizations are drawn from an 
i.i.d. $\mathcal{N}(0,\sigma^2)$ (zero mean white Gaussian noise) process, 
with $\sigma = 100$~ns for all pulsars. 
The higher observational frequency band of SKA may also improve data quality by
mitigating the problem of red noise \cite{2015Sci...349.1522S}.
\paragraph{Optimal Signal to Noise Ratio --} It is convenient to characterize 
a PTA using its
network signal-to-noise ratio (SNR), $\rho$, defined as
\begin{eqnarray}
\rho & = & \left[ \sum_{I = 1}^{N} \rho_I^2\right]^{1/2}\;,\\
\rho_I & = & \| F_+^I(\alpha,\delta)\Delta h_+ + F_\times^I(\alpha,\delta)\Delta h_\times\|/\sigma_I\;.
\end{eqnarray}
Here, 
$\rho_I$ is the individual optimal SNR for the $I^{\rm th}$ pulsar, $\sigma_I = \sigma$ in our simulations, and
$\| v \|^2 =\sum_{i=1}^{k} v_i^{2}$ for $v \in \mathbb{R}^k$. 

\begin{figure}
\centerline{\includegraphics[scale=0.48]{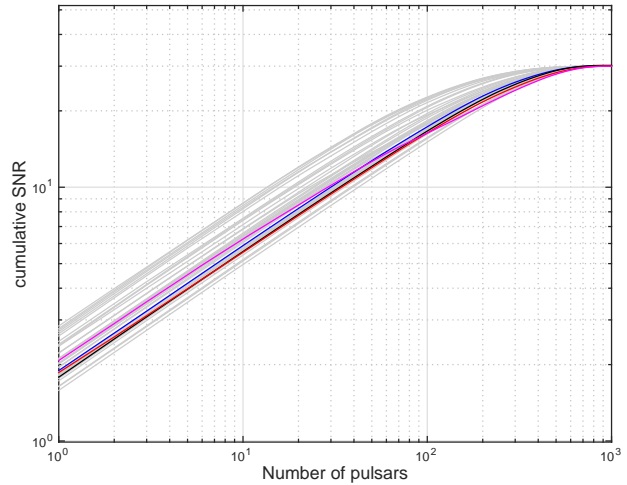}}
\caption{Cumulative network SNR for a uniformly spaced 6-by-6 grid of source 
locations (grey) and the four locations, A (blue), B (black), 
C (red), and D (magenta), used in the simulations. 
The total network SNR, $\rho$, is $30$ for this plot. The curve for any other $\rho$ 
can be obtained by using an overall scale factor of $\rho/30$. 
\label{fig:accSNR3_loc6789}
}
\end{figure}

Fig.~\ref{fig:accSNR3_loc6789} shows
the cumulative network SNR for a representative SMBHB system 
when the $\rho_I$ are arranged in descending order.  It can be seen that 
a substantial fraction of pulsars must be included 
in order to avoid a significant loss of network SNR. 
Almost independently of the 
source location, contributions from $> 200$
pulsars are needed to reach 90\% of the total SNR. 
Taking only the top 20 pulsars, one gets less than 40\% of the total SNR. 
Non-uniformly distributed noise levels ($\sigma_I$) in a real PTA
will enhance the location dependence of the
required fraction of pulsars but, qualitatively, give the same result. 

It should be noted that while the metric $\rho$ is simple to compute, 
it pertains to the best-case scenario for a search where the GW signal parameters are 
known {\it a priori}. In reality, detection 
requires the global maximum, over all the signal parameters, of the joint 
log-likelihood function of the full data from a PTA.
The 
resulting effect of
the parameter space volume on the false alarm probability of the detection statistic 
is not accounted for in $\rho$. 


\paragraph{Pulsar phase parameters --}
For the large fraction of  SMBHB sources that are  expected to 
evolve slowly~\cite{2015MNRAS.451.2417R}, $h_{+,\times}(t)$ is approximately monochromatic. 
The time delay $\tau_I(\alpha,\delta)$ (c.f. Eq~\ref{eq:pulsarterm}) then transforms into a 
fixed phase offset, $\varphi_I$, called the pulsar phase parameter. 
Uncertainty in our knowledge of the Earth-pulsar distance 
makes $\varphi_I$ an {\em a priori} unknown quantity even if $\alpha$ and $\delta$
are known. 
Hence, every pulsar in a PTA contributes a new parameter 
to the joint log-likelihood. For a SKA era PTA, this 
leads to an 
infeasible
optimization problem over hundreds of unknown
parameters since, as discussed earlier,  the bulk of the pulsars must be included in
a search.

A solution to the pulsar phase problem is 
provided by a judicious choice of the parameters
that are maximized over (semi-)analytically 
in the optimization process.
Choosing the pulsar phase parameters as this subset \cite{2014ApJ...795...96W} leads to the MaxPhase algorithm~\citep{2015ApJ...815..125W}. 
The remaining optimization, 
involving a fixed
7-dimensional search space,
is carried out using Particle Swarm 
Optimization (PSO) \cite{eberhart1995new, 2010PhRvD..81f3002W, 2014ApJ...795...96W}. 
(The PSO algorithm used here is slightly modified  
to improve performance for angular variables.) 
The applicability of the alternative approach of numerically optimizing over the 
pulsar phase parameters \cite{2016MNRAS.461.1317Z} 
has not been established for $\gtrsim 30$ pulsars. 
A method~\cite{2014PhRvD..90j4028T} that obtains a 7 dimensional search space
by marginalizing over the pulsar phases has been applied to 41 MSPs in~\cite{2016MNRAS.455.1665B}.


\paragraph{Results --}
We assess the detection and estimation performance of MaxPhase for the simulated PTA 
in the context of 
(i) an all-sky search, 
with unknown source location, 
and (ii) known candidate SMBHB systems. For the latter,
we take PG~1302-102 \cite{2015Natur.518...74G} 
and PSO~J334+01 \cite{2015ApJ...803L..16L} as examples. 
While PSO~J334+01 may be near coalescence by the time SKA starts (around 2025), it serves as a prototype for
similar candidates that may be found when the Large Synoptic Survey Telescope (LSST)\footnote{https://www.lsst.org}
begins operation on roughly the same timescale (around 2023). 

In order to quantify the effect of ill-posedness discussed earlier, we pick 
simulated source locations as shown  in Fig.~\ref{fig:Acond},
that correspond to a range of condition numbers. These source locations, denoted
as A, B, C, and D, have DEC (in radians) of $0.3$, $-0.2$, $-0.7$, and $-1.2$ 
respectively but the same RA of $3.5$~radian.

Besides source location, 
the parameters defining a SMBHB GW signal consist of the observer frame quantities 
$\zeta$ (the overall timing residual amplitude), $f_{\rm gw}$ (GW signal frequency), $\iota$ (the inclination angle), 
$\psi$ (polarization angle), and $\varphi_0$ (the initial orbital phase of the binary).  
We scale $\zeta$ to get the desired network SNR $\rho$ and keep identical values for the
remaining parameters across the four simulated sources: $f_{\rm {gw}} = 2\times 10^{-8}$~Hz, 
$\iota = 0.5$, $\psi = 0.5$ and $\varphi_0=2.89$.
 
Consider a subset of the SKA era PTA with $\sim 30$ pulsars, 
the maximum that methods based on numerical optimization over 
pulsar phase parameters can handle at present \citep{2016MNRAS.461.1317Z}. 
Assuming a marginal detection network SNR $\rho \simeq 10$ for such a subset, 
we see from Fig.~\ref{fig:accSNR3_loc6789} that the same source will have 
$\rho \simeq 30$ for the full PTA. 
We set this as the fiducial value for the discussion of 
detection performance below. 

As discussed earlier, $\rho$ alone does not quantify the actual performance of a 
detection statistic. To make a proper assessment, 
simulations were carried out with 200 realizations of data 
containing only noise, and 50 realizations for each source location 
containing signal plus noise.  We find that the distributions of the 
MaxPhase statistic are fit well by (i) 
a log-Normal distribution 
${\rm ln}\,\mathcal{N}(6.44, 3.80\times 10^{-4}) $ for the noise-only case, and (ii)
by Normal distributions for all the simulated sources. 
For a 
conservative estimate of detection probability, we pick the Normal distribution with the lowest mean value
($\mathcal{N}(1067.25, 2045.82)$). From these fits, 
the detection probability is $99.99\%$ at a false alarm probability of $ 10^{-4}$.

Having established that $\rho = 30$ corresponds to a high confidence detection, we use
the relations given below to translate $\rho$ into quantities of astrophysical interest.
\begin{eqnarray}
\label{eq:amp}
\zeta &=& 4.8 \times 10^{-10} \left(\frac{\mathcal{M}_c}{10^9~M_\odot}\right)^{5/3} 
\left(\frac{D}{7.2~\text{Gpc}}\right)^{-1}   \nonumber \\
&& \times \left(\frac{f_{\text{gw}}}{2\times10^{-8}~\text{Hz}}\right)^{-1/3} \text{sec}  \,, \\
\label{eq:rho2zeta}
\left(\frac{\rho}{30}\right) & = &\kappa\, \mathcal{G}(\alpha,\delta)  \left(\frac{\zeta}{5.1\times 10^{-10}}\right)  \;,\\
\label{eq:rho2hG}
\left(\frac{\rho}{30}\right)&=&\kappa\, \mathcal{G}(\alpha,\delta) \left(\frac{h}{6.4\times 10^{-17}}\right)
\left(\frac{f_{\text{gw}}}{2\times10^{-8}~\text{Hz}}\right)^{-1}    
\end{eqnarray}
where, $\kappa = \left(T/5\,\text{yr}\right)^{1/2} 
\left(\sigma/100\,\text{ns}\right)^{-1}$.
Here, (i) $\mathcal{M}_c = (1+z)M_c$ is the {\em observed} (redshifted) chirp
mass, with $M_c=(m_1 m_2)^{3/5}/(m_1 + m_2)^{1/5}$ being the 
chirp mass in the rest frame of a source having component masses $m_1$ and $m_2$, (ii)
$D$ is the luminosity distance 
(related to redshift $z$ through standard values of cosmological parameters 
\cite{2015arXiv150201589P}), (iii) $T$ is the observation span, (iv) $h$ is the overall 
GW strain amplitude, and (v) $\mathcal{G}$ is a geometrical
factor that arises, after averaging 
over $\psi$ and $\iota$, from the antenna pattern functions and ranges over 
$[0.87, 1.6]$ for the simulated PTA, with a sky-averaged value of $1.2$.

For $\rho = 30$, a SMBHB with the fiducial parameters used in Eq.~\ref{eq:amp} 
will be detectable in a redshift
range, corresponding to the variation in $\mathcal{G}$, of $[0.95,1.55]$. 
A system with $\mathcal{M}_c = 10^{10}$~$M_\odot$, on the other hand, will be visible 
with this $\rho$ out to $z = 28.03$, which is well beyond $z \simeq 2$
where the SMBHB formation rate is thought to peak \cite{2003ApJ...582..559V, 2013ApJ...773...14R}.

In the absence of a detection, an upper limit can be set on the GW strain 
amplitude averaged across the sky. If $\rho = 30$ is used as a detection
threshold, a non-detection can rule out a GW strain of  
$\geq 5.2 \times10^{-17}$ at  $f_{\text{gw}}=2\times10^{-8}$ Hz,
a significant improvement over the most 
stringent PTA based upper limit for continuous waves to date 
($\approx 10^{-14}$ at  $f_{\text{gw}}=2\times10^{-8}$ Hz) \citep{2016MNRAS.455.1665B}.

Table~\ref{table:candidateParams} lists the relevant parameters 
 obtained from optical observations of the candidate systems. 
Based on these values and Eq.~\ref{eq:rho2hG} (with $\mathcal{G}$ set to its sky-averaged value), 
the predicted GW strain amplitudes
range over $(6 \times 10^{-18}, 4 \times 10^{-16})$ for PG~1302-102 and
$(6 \times 10^{-16}, 2 \times 10^{-15})$ for PSO~J334+01 corresponding to their respective uncertainties in 
redshifted chirp mass. These 
are well below the upper limits, $\gtrsim 1.0 \times 10^{-14}$ set by 
current PTAs \cite{2014MNRAS.444.3709Z} at the respective GW emission
frequencies (twice the orbital frequencies) of these systems. However, these
are well within the reach of a SKA era PTA.
\begin{table}
\begin{center}
\begin{tabular}{c|c|c|c|c|c}
\hline\hline
Candidate & $\alpha$ (rad) & $\delta$ (rad) & $P$ (yr) &$\mathcal{M}_c$ ($M_\odot$) & $z$ \\
\hline
PG~1302-102 & 3.4252 & -0.1841 & 5.2 & $10^{8.0}-10^{9.1}$ & 0.2784\\
PSO~J334+01  & 0.9338 & 0.0246  & 1.48 & $10^{9.6}-10^{10.0}$ & 2.06\\
\hline
\end{tabular}
\end{center}
\caption{Relevant parameters of the candidate SMBHB systems considered in this letter. 
$P$ denotes the observed orbital period.}
\label{table:candidateParams}
\end{table}

A non-detection of the GW signal at $\rho \geq 30$ from 
PG~1302-102 with a SKA era PTA will rule out, with very high confidence, 
a value of $\mathcal{M}_c \geq 10^{8.67}$~${\rm M}_\odot$. 
The corresponding upper limit on the rest frame total mass is $\leq 10^{9.01}$~${\rm M}_\odot$. 
For PSO~J334+01, the signal will have $\rho>100$ regardless 
of the uncertainty in the chirp mass. 
MaxPhase is sub-optimal for evolving signals, 
but any reasonably sub-optimal algorithm can confidently detect such a strong signal.
Therefore, such a system should be a {\it guaranteed} source for a SKA era PTA. 

The location of the global maximum of the log-likelihood provides
the Maximum Likelihood point estimate for the GW signal parameters. 
To study the dependence of the 
estimation errors on signal strength, we carried out additional simulations for $\rho = 60$ and $100$, 
with 50 data realizations for each of the locations A, B, C and D. As expected, the 
frequency $f_{\rm gw}$ is the best estimated parameter with a standard deviation ranging from $\sim 1\%$
to $\sim 0.1\%$ (relative to the estimated mean) for the lowest ($\rho = 30$) to the highest value of $\rho$ respectively.
The corresponding range for the parameter $\zeta$, are $[11.0\%, 7.5\%]$, respectively. 
The estimates of $\iota$, $\psi$, and $\varphi_0$ show a non-negligible bias while
their standard deviation typically ranges over a few ten percents. 
In the following,  we focus on  the localization of a SMBHB source on the sky.

\begin{figure*}
\centerline{\includegraphics[width=1.0\textwidth]{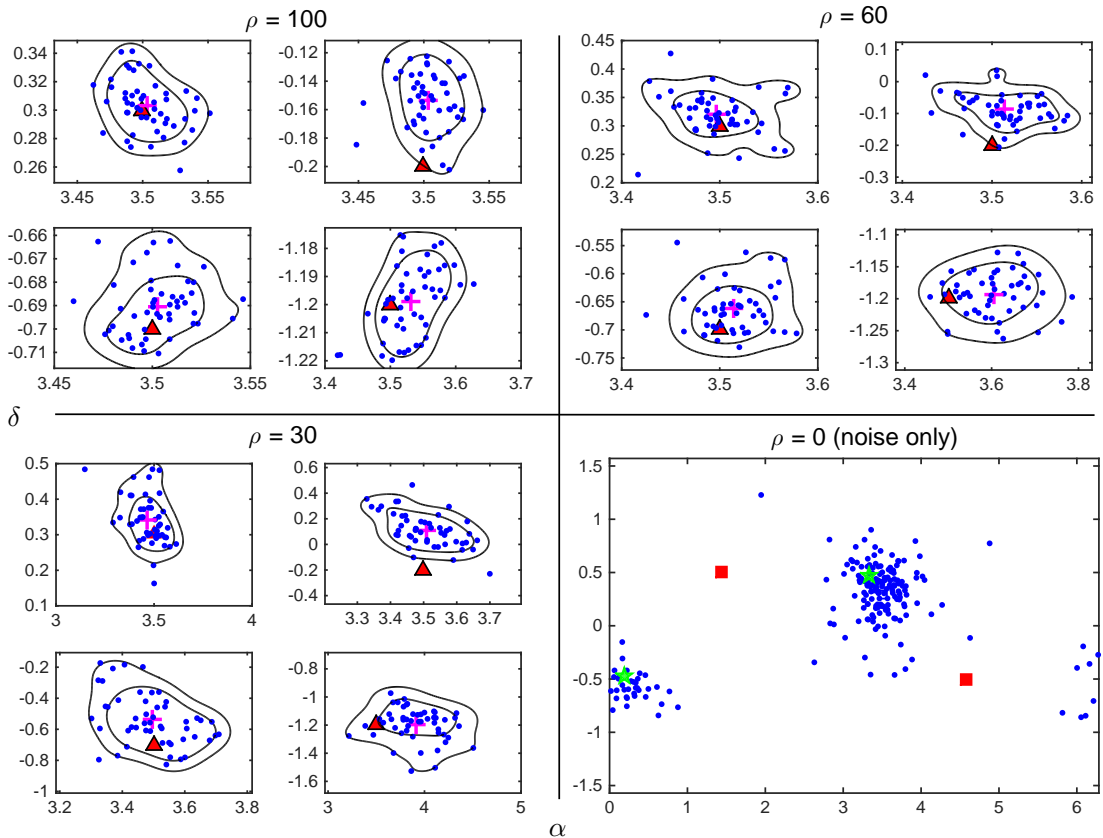}}
\caption{Maximum likelihood estimates (blue dots) of the GW source location in equatorial coordinates with
50 data realizations for the network SNR $\rho \neq 0$ and 200 for $\rho = 0$ (noise only). 
The subpanels for each $\rho \neq 0$ panel correspond to the true locations A (top left), 
B (top right), C (bottom left) and D (bottom right), respectively. 
The true and estimated mean locations are marked by red triangles and magenta crosses 
respectively. The solid black lines show regions in which the probability of getting an
estimated location is 68\% and 
95\% (estimated using Kernel Density Estimation \cite{botev2010}). 
In the $\rho = 0$ panel, the center and anti-center of the Galaxy and its poles
 are marked by red squares and green stars respectively.  }
\label{fig:loc_loc6789}
\end{figure*}

Fig.~\ref{fig:loc_loc6789} shows the estimated 
sky positions for the different values of $\rho$ and source locations used in our simulations.  
The condition number of the antenna pattern matrix $\textbf{A}$ is seen to have 
an important effect on both estimation bias and variance. 
It affects the noise only case ($\rho=0$) by concentrating the estimated locations 
around the two galactic poles where its value approaches
unity. These two locations also act as  attractors 
when $\rho > 0$ by introducing a bias in the estimation.
This is most clearly seen for locations B and D where the estimates are attracted towards
the galactic north and south poles respectively. However,
except for location B, the true locations fall within the 95\% confidence regions 
associated with the estimates. 

At $\rho = 30$, and excluding location B, the standard deviations 
$\sigma_\alpha$ and $\sigma_\delta$ 
of $\alpha$ and $\delta$ respectively are 
$\sigma_\alpha = (4.76^{\circ}, 
6.25^{\circ}, 
15.2^{\circ})$,
and 
$\sigma_\delta = (
3.90^{\circ}, 
9.57^{\circ}, 6.70^{\circ} 
)$
for Loc A, 
C and D respectively. Making the conservative but simple choice of 
$(2 \sigma_\alpha) (2\sigma_\delta) \cos\delta $ as the error area, 
the sources can be localized to within $\sim 70$ to $\sim 180$~${\rm deg}^2$. 
As demonstrated in the search for 
PSO~J334 \citep{2015ApJ...803L..16L}, 
which used a $\sim 80$~${\rm deg}^2$ field from the Pan-STARRS1 Medium Deep Survey,
this may be accurate enough to permit 
host galaxy identification in optical follow-ups. 
The joint operation of SKA with LSST will further boost the prospects of such multi-messenger studies of SMBHBs.

\paragraph{Limitations of the study and future work --}
As is common in studies of isolated 
sources~\cite{2012ApJ...756..175E, 2015MNRAS.449.1650Z}, 
the signal from unresolved SMBHBs was ignored under the implicit assumption that 
it simply elevates the noise level. Future studies should test this assumption. 

The fixed observed frequency of the signal in our
simulation translates at a sufficiently high redshift into a rest frame frequency corresponding
to a rapidly evolving phase of the binary \cite{2016PhRvL.116j1102R}. 
However, for the redshifted chirp
masses considered here, this effect will only manifest itself at redshifts $z \gg 2$, the epoch of 
peak SMBHB formation rate, and may be ignored. 

In the future, we plan to incorporate some form of regularization 
\cite{2006CQGra..23S.673R, 2006CQGra..23.4799M} in MaxPhase 
to mitigate the adverse effects of ill-posedness seen on source localization. 
The algorithm will be refined further by taking uncertainties in the measured
noise parameters \citep{2016ApJ...821...13A} into account.
Additionally, it will be extended to include non-monochromatic signal models.

\begin{acknowledgments}
We acknowledge Roy Smits at the Netherlands Institute 
for Radio Astronomy (ASTRON) for providing us the SKA pulsar simulation. 
Y.W. is supported by the National Natural Science Foundation of China 
under grants 11503007, 91636111 and 11690021. 
The contribution of S.D.M. to this paper is supported by NSF awards PHY-1505861 and HRD-0734800. 
We thank the anonymous referees for helpful comments and suggestions. 
\end{acknowledgments}


\end{document}